\documentclass[doublecol]{epl2}

\usepackage{amsfonts}
\usepackage{amsthm}
\usepackage{amsbsy}
\usepackage{amssymb}
\usepackage{amsmath}
\usepackage{graphics}
\usepackage{color}

\usepackage{epsfig}

\newcommand{\rr}{\mathbf{r}}
\newcommand{\ah}{\hat{a}}
\newcommand{\ahd}{\hat{a}^\dagger}
\newcommand{\eqname}[1]{\label{eq:#1}}
\newcommand{\eq}[1]{(\ref{eq:#1})}

\title{Dynamical Casimir Effect in Optically Modulated Cavities} 

\author{Daniele Faccio\inst{1}\thanks{E-mail:
\email{d.faccio@hw.ac.uk}}, Iacopo Carusotto\inst{2}} 
\shortauthor{D. Faccio,  I. Carusotto}

\institute{ \inst{1} School of Engineering and Physical Sciences, SUPA, Heriot-Watt University, Edinburgh EH14 4AS, United Kingdom \\ 
\inst{2} INO-CNR BEC Center and Dipartimento di Fisica, Universit\`a di Trento, via Sommarive 14, 38123 Povo-Trento, Italy}


\abstract{
Cavities with periodically oscillating mirrors have been predicted to excite photon pairs out of the quantum vacuum in a process known as the Dynamical Casimir effect. Here we propose and analyse an experimental layout that can provide an efficient modulation of the effective optical length of a cavity mode in the near-infrared spectral region.
An analytical model of the dynamical Casimir emission is developed and compared to the predictions of a direct numerical solution of Maxwell's equations in real time. A sizeable intensity of dynamical Casimir emission is anticipated for realistic operating parameters. In the presence of an external coherent seed beam, we predict amplification of the seed beam and the appearance of an additional phase-conjugate beam as a consequence of stimulated dynamical Casimir processes.
}

\begin{document}

\maketitle 

\section{Introduction}
On the basis of quantum field theory, a neutral planar mirror moving with a non-uniform acceleration is expected to convert the zero point fluctuations of the quantum vacuum of the electromagnetic field into real propagating photons\cite{fulling}. This so-called Dynamical Casimir Effect (DCE), predicted more than 30 years ago, has proved extremely hard to experimentally observe due to the extreme weakness of the emitted photon flux for realistic configurations.
Resonant enhancement of the DCE intensity inside an optical cavity with oscillating mirrors was investigated~\cite{lam1,lam2,crocce,uhl}, still a sizable DCE emission requires a relativistic motion of the mirror, which is almost impossible to obtain using material mirrors in real mechanic motion.

To overcome this difficulty, alternative methods have been explored, based on the modulation of the effective {\em optical length} of the cavity. A scheme based on the modulation of the skin depth of a semiconductor mirror was proposed in \cite{yablo} and is presently being experimentally implemented\cite{braggio}. Schemes based on coupling the cavity mode to an emitter with time-dependent properties were explored in~\cite{noi1,noi2}.

The recent observation of a microwave emission from a modulated SQUID into a superconducting circuit has been considered as a first experimental evidence of DCE~\cite{delsing}: as the phase of microwave reflection on the SQUID depends on the applied magnetic field, a spatially moving mirror can be simulated by rapidly varying the magnetic field imposed to the SQUID, which has been predicted to lead to an appreciable DCE emission~\cite{johansson}.

In the present paper, we theoretically explore a strategy to observe the DCE in the optical domain by modulating in time the refractive index of the medium filling the cavity. This strategy to modulate the optical length of the cavity, originally proposed in~\cite{DCrefr,law,artoni}, has recently been pushed further in~\cite{new_DCrefr} and bears a tight resemblance to the concept of time refraction \cite{mendonca2} and to recent efforts at using optical analogues for the study of cosmological particle creation and black hole evaporation \cite{philbin,belgiorno}. 

Even though our theoretical developments are very general, here we focus our attention on a specific scheme where the optical length of the cavity is modulated in time by including in the cavity a slab of $\chi^{(3)}$ nonlinear optical material and pumping it with a periodic train of ultrashort optical pulses with an (angular) repetition rate $\Omega=2\pi/T$, $T$ being the time separation of neighbouring pulses. As a result, the cavity mode (of unperturbed frequency $\omega_c$) experiences an effective refractive index of the nonlinear slab which follows the instantaneous pump intensity $n_{\rm eff}=n_0+\delta n(t)=n_0+n_2\,I_p(t)$. As usual, the DCE emission is maximum when the resonance condition $\Omega\simeq 2\omega_c$ is satisfied.

To quantitatively describe the DCE emission, a quantum optical model of the modulated cavity is developed, which can be analytically solved and where the details of the specific configuration are summarized by a few parameters.
Inserting in the model realistic values of the  parameters for a pump pulse train resulting from the interference of several harmonics of a fundamental laser beam at 1064 nm, an experimentally sizeable DCE emission in the near-IR around a wavelength of $2\,\mu$m is predicted. Using the same model,  we also investigate the response of the cavity to a seed beam which is incident on the cavity at a frequency $\omega_s$ close to resonance with the cavity mode:  in the presence of the modulation of the refractive index at $\Omega$, the stimulated counterpart of the DCE results in the amplification of the seed beam after being transmitted through the cavity, and in the appearance of a phase-conjugated beam at a frequency $\Omega-\omega_s$ due to the scattering of the seed beam on the time-dependent refractive index. Our predictions are finally verified by numerically solving Maxwell's equations for a cavity filled with a linear medium with a time-dependent refractive index.

\section{The physical system}

\begin{figure}[!t]
\centering
\includegraphics[width=8cm]{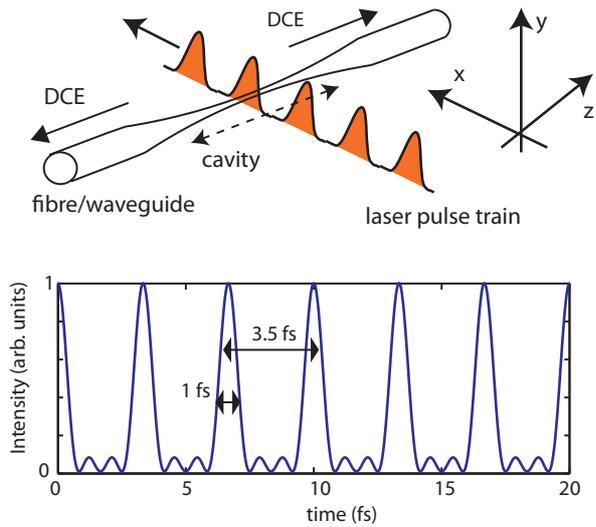}
\caption{\label{fig:layout}  Upper panel: Proposed experimental scheme for studying the DCE. The optical cavity resides within a photonic nanowire or waveguide and is externally excited by a periodic laser pulse train that modulates the cavity optical length through the nonlinear Kerr effect. DCE photons are excited from the vacuum state and transmitted along the waveguide which terminate with photon detectors. In addition,  the cavity may be seeded by a optical signal sent down the waveguide from one end and detected at the other end.
Lower panel: simulated pulse train obtained by superimposing the first four harmonics of a nanosecond Nd:YaG laser at a fundamental wavelength of 1.064 $\mu$m.}
\end{figure}

The system we are studying is based on an optical waveguide cavity as shown in the schematic drawing in Fig.~\ref{fig:layout}. The cavity is embedded within an optical waveguide, e.g. as a Bragg grating cavity. 
A train of ultrashort, intense laser pulses separated by a time interval $T$ is sent externally on to the waveguide, ideally at 90 deg. The laser pulses, coupled to the medium through the $\chi^{(3)}$ optical nonlinearity of the cavity material, induce a periodic modulation of the effective refractive index $n_{\rm eff}(t)=n_0+\delta n(t)=n_0+n_2\,I_p(t)$, where $n_0$ is the refractive index of the unperturbed cavity mode, $n_2$ is the nonlinear Kerr coefficient (proportional to the $\chi^{(3)}$ nonlinear susceptibility) and $I_p(t)$ is the instantaneous intensity of the pulse train. 

Using the typical value $n_2=3\cdot 10^{-16}$ cm$^2$/W for fused silica, pulses with a reasonable peak intensity on the order of  $10^{13}$ W/cm$^2$ will modulate the cavity refractive index between $n_0$ and $n_0+0.003$. A pulse train with the required intensities can be generated  by combining the first four harmonics of a nanosecond Nd:YaG laser in a similar fashion to pulse train generation techniques already demonstrated in the literature \cite{harris,kung}. 
As shown in  the lower panel of Fig.~\ref{fig:layout}, this will create a train of pulses with sub-femtosecond duration and spatially separated by the fundamental wavelength $\Lambda=1064$~nm of the beam, which gives an (angular) repetition rate  $\Omega=2\pi c/\Lambda=(2\pi)\,0.3$~PHz. Less than 10 mJ energy, equally distributed between the four harmonics, is sufficient to guarantee the required intensities. Care should be taken that the cavity transverse dimension is kept smaller than the distance between each pulse so as to ensure that only one pulse at a time modulates the cavity length. Typical ridge waveguides, photonic crystal cavities or photonic nanowires all satisfy this requirement and are viable solutions. The time-modulation of the refractive index of the cavity will excite real $2$ $\mu$m photons out of the vacuum state by DCE, which can be efficiently collected at the output ports of the waveguide and detected with state of the art single photon detectors.

\section{The theoretical model}

For the sake of simplicity, we restrict our description to a single cavity mode with an electric field profile
\begin{equation}
\mathcal{E}_c=\mathcal{E}_c^0\cos(k_c z)\,e^{-(x^2+y^2)/2\sigma_c^2}.
\end{equation} 
fully contained within the fiber.
The $z$-axis is oriented along the cavity axis and $\sigma_c$ is the mode waist in the transverse $x$ and $y$ directions. For a cavity of length $L_c$ filled with a material of dielectric constant $\epsilon_c$, the single-photon amplitude is equal to $\mathcal{E}_c^0=\sqrt{4\hbar\omega_c/L_c\sigma_c^2\epsilon_{c}}$. 

The pump beam consists of a train of optical pulses of duration $\tau$ separated by a much longer time interval $T\gg \tau$, that propagate across the cavity at a speed $v_p$ along the direction $x$, orthogonal to the cavity axis.  The peak electric field amplitude is $E_p^0$. Along $y$ and $z$, the pump has a transverse Gaussian profile of wide waist $\sigma_p\gg 1/k_c, \sigma_c$. In formulas, this corresponds to an electric field profile $E_p(\rr,t)=\bar{E}_p(\rr,t)\,e^{i(k_px-\omega_pt)}$ with carrier frequency $\omega_p$, carrier wavevector $k_p$, and envelope 
\begin{equation}
\bar{E}_p(\rr,t)=\sum_n E_p^0\,e^{-[x-v_p (t-nT)]^2/2\sigma^2}\,\\ 
e^{-(y^2+z^2)/2\sigma_p^2}.
\end{equation}
Here, the sum over $n$ runs over the pulses forming the train.  $\sigma=v_p\tau$ is the spatial length of each pulse within the fibre material, which is assumed to be non-dispersive.

The Hamiltonian describing the dynamics of the single cavity mode under the nonlinear modulation $\delta \epsilon(\rr,t)$ of the dielectric constant induced by the pump train of pulses has the form~\cite{belgiorno_pulse}
\begin{equation} 
\eqname{H_eff}
\delta H=-\!\!\int \!\!d^3\rr\,\frac{\delta \epsilon(\rr,t)\,[\mathcal{E}_c(\rr)]^2}{8\pi}\left[\ah_c + \ahd_c\right]^2=\mathcal{A}(t)\left[\ah_c + \ahd_c\right]^2
\end{equation}
where $\ah_c$ and $\ahd_c$ are the destruction and creation operators for photons in the cavity mode. 
The time dependence of the effective coupling constant 
\begin{equation}
\mathcal{A}(t)=\mathcal{A}_0\sum_n e^{-(t-nT)^2/\bar{\tau}^2}
\end{equation}
then consists of a train of peaks of height 
\begin{equation}
\mathcal{A}_0=
-\frac{\sqrt{\pi}}{2}\,\frac{\delta n^{\rm peak}}{n_0}\,\hbar\omega_c\,\frac{\sigma_p}{L_c}\,
\frac{\sigma}{v_p \bar{\tau}}
\eqname{A0}
\end{equation}
and effective duration 
\begin{equation}
\bar{\tau}=\sqrt{\sigma_c^2+\sigma^2}/v_p
\end{equation}
that result from the spatial overlap of the cavity mode with the pump beam. Note that the effective duration $\bar{\tau}$ can be significantly longer than the pulse duration $\tau$ as soon as the waist $\sigma_c$ of the cavity mode exceeds the pulse length $\sigma=v_p\tau$.
The peak modulation of the refraction index is 
\begin{equation}
\delta n^{\rm peak}=n_2 I_p^{\rm peak}= \frac{ c n_2}{2\pi n_0}\,|E_p^0|^2.
\end{equation}

The Fourier transform of $\mathcal{A}(t)$ consists of a comb of $\delta$-peaks spaced by the repetition rate $2\pi/T$ and multiplied by a broad Gaussian envelope of width proportional to the inverse pulse duration $\bar{\tau}^{-1}$,
\begin{equation}
\tilde{\mathcal{A}}(\omega)=\frac{2\pi^{3/2}\bar{\tau}\mathcal{A}_0}{T}\sum_{j=-\infty}^{\infty}\delta\left(\omega-\frac{2\pi}{T}j\right)\,e^{-\omega^2\,\bar{\tau}^2/4}.
\eqname{Atilde}
\end{equation}

In the following we shall neglect the small nonlinear frequency shift of the cavity mode due to the $\ahd_c\ah_c$ terms in \eq{H_eff} and concentrate our attention on the processes where two cavity photons are either created or destroyed by Hamiltonian terms proportional to $(\ahd_c)^2$ or $\ah_c^2$. These processes are strongest when one of the components of $\tilde{\mathcal{A}}(\omega)$ is close to resonance with twice the cavity mode frequency. From now on, we shall assume this condition to be approximately met for the $j$-th component at $\bar{\Omega}=j\Omega$, and we neglect all other components,
\begin{equation}
\mathcal{A}(t)\simeq \frac{\sqrt{\pi}\mathcal{A}_0 \bar{\tau}}{T}\,e^{-\bar{\Omega}^2\,\bar{\tau}^2/4}\,e^{-i\bar{\Omega} t}+\textrm{c.c.}=\hbar \mathcal{B}e^{-i\bar{\Omega} t}+\textrm{c.c.}
\eqname{Afin}
\end{equation}
This leads to the final form for the isolated cavity Hamiltonian,
\begin{equation}
H_0=\hbar \omega_c \ahd_c \ah_c+ \hbar \mathcal{B}\,\left( e^{-i\bar{\Omega} t} \ah^{\dagger2}_c + e^{i\bar{\Omega} t} \ah_c^2\right)
\eqname{H0}
\end{equation}
For strong enough modulations $2\mathcal{B}>\bar{\Omega}/2-\omega_c$, the Hamiltonian \eq{H0} for the isolated, lossless cavity  predicts that  the number of cavity photons exponentially grows in time at a rate
\begin{equation}\label{gamma}
\Gamma=2\sqrt{4\mathcal{B}^2-\left(\frac{\bar{\Omega}}{2}-\omega_c\right)^2};
\end{equation}
this exponential amplification of the cavity field will be recovered in the section about numerical calculations using an {\em ab initio} FDTD simulation of the electromagnetic field under a time-modulation of the dielectric constant.

To go beyond this very idealized model and be able to describe the steady state of realistic cavities, one has to include radiative and non-radiative losses and, possibly,  the external driving of the cavity by means of incident light beams. In standard treatments~\cite{wallsmilburn,quantumnoise}, the Hamiltonian for a single cavity mode coupled to external light sources is written in the form
\begin{equation}
H=H_0 + \hbar \kappa_c \left[E_{\rm inc}(t) \,\ahd_c +E_{\rm inc}^*(t) \,\ah_c\right],
\end{equation}
where $E_{\rm inc}(t)$ is the amplitude of the coherent laser field incident on the cavity and $\kappa_c$ is a coefficient quantifying the coupling of the incident radiation to the cavity mode. 
Losses are then included at the level of the master equation for the density operator
\begin{equation}
\partial_t\rho=-\frac{i}{\hbar}[H,\rho]+\frac{\gamma_c}{2}\{2\ah_c\rho\ahd_c-\rho\ahd_c\ah_c -\ahd_c\ah_c\rho\},
\end{equation}
where $\gamma_c$ is the total cavity decay rate.

\section{Analytical predictions}

Standard quantum optical techniques such as the input-output formalism or the semi-classical Wigner representation of the quantum field $\ah_c$ can be used to obtain exact predictions for the most significant observables in the steady-state reached by the cavity under the interplay of the Casimir modulation, the losses, and possibly a coherent incident laser beam~\cite{wallsmilburn,quantumnoise}.
In the absence of a coherent drive $E_{\rm inc}=0$, the total photon emission rate 
\begin{equation}
\eqname{N}
\Phi_{\rm DCE}=\frac{2\gamma_c \mathcal{B}^2 }{\gamma_c^2/4-4\mathcal{B}^2+(\omega_c-\bar{\Omega}/2)^2}
\end{equation}
vanishes in the absence of any modulation $\mathcal{B}=0$ and is strongest at the DCE resonance, i.e. when the modulation frequency $\bar{\Omega}$ is close to twice the cavity frequency $\omega_c$. The divergence of the emission rate for $\mathcal{B}\rightarrow \mathcal{B}_{\rm thr}$ with 
\begin{equation}
\mathcal{B}_{\rm thr}=\frac{1}{2} \sqrt{\frac{\gamma_c^2}{4} + \left(\omega_c-\frac{\bar{\Omega}}{2}\right)^2}
\end{equation} signals the threshold for coherent oscillation in the cavity by a mechanism which is the dynamical Casimir analog of parametric oscillation.  A related parametric oscillation effect via DCE was observed in~\cite{delsingOPO} using a superconducting circuit cavity whose electric length is modulated in a fast and periodic way by an external magnetic field driving the terminating SQUID.

\begin{figure}[!t]
\centering
\includegraphics[width=6.5cm,angle=270]{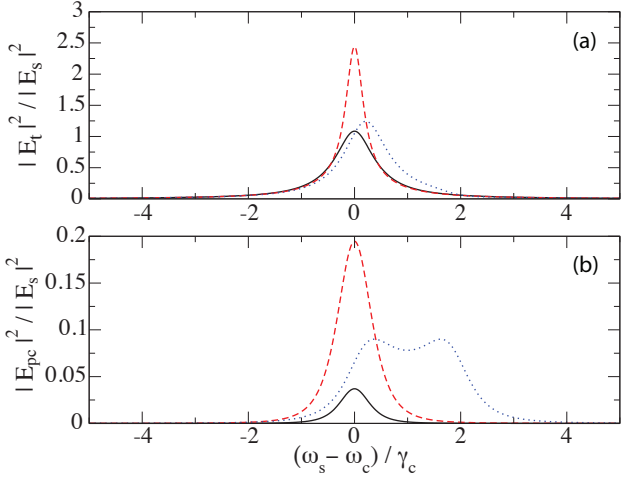}
\caption{\label{fig:stimulated} 
Relative intensity of the transmitted [panel (a)] and phase-conjugated [panel (b)] beams as a function of the seed  frequency $\omega_s$. Black solid lines: $\bar{\Omega}=2\omega_c$, $\mathcal{B}/\gamma_c=0.05$. Red dashed lines: $\bar{\Omega}=2\omega_c$, $\mathcal{B}/\gamma_c=0.15$. Blue dotted llines: $\bar{\Omega}-2\omega_c=2\gamma_c$, $\mathcal{B}/\gamma_c=0.3$. }
\end{figure}

In the presence of a coherent seed $E_{\rm inc}(t)=E_s^0\,e^{-i\omega_s t}$  at frequency $\omega_s$, the coherent emission on the opposite side of the cavity consists of two monochromatic beams: a transmitted beam  at the seed frequency $\omega_s$ and a phase-conjugated beam  at the frequency $\bar{\Omega}-\omega_s$, specular with respect to $\bar{\Omega}/2$. For a symmetric cavity, the two emerging beams have amplitudes 
\begin{eqnarray}
E_{ t}=\frac{\gamma_c/2}{ \omega_s-\omega_c+i\frac{\gamma_c}{2} + \frac{4\mathcal{B}^2}{\omega_s+\omega_c-\bar{\Omega}+i\frac{\gamma_c}{2}}}  E_s \\
E_{pc} = \frac{-\mathcal{B}\gamma_c}{(\omega_s-\omega_c+i\frac{\gamma_c}{2})(\omega_s+\omega_c-\bar{\Omega}+i\frac{\gamma_c}{2})+4\mathcal{B}^2}\,E_s^* 
\end{eqnarray}
Examples of the $\omega_s$-dependence of their intensities are plotted in Fig.~\ref{fig:stimulated} for different values of the modulation amplitude $\mathcal{B}$ and of the detuning $\bar{\Omega}-2\omega_c$.

In the absence of modulation $\mathcal{B}=0$, the amplitude $E_t$ of the first component [Fig.~\ref{fig:stimulated}(a)] reduces to standard resonant transmission through the cavity: it is complete on resonance $\omega_s=\omega_c$ and the resonance peak has a linewidth $\gamma_c$. 
For finite modulations $\mathcal{B}$, the transmitted intensity can grow above one, which signals the onset of stimulated DCE processes. For increasing values of the modulation amplitude $\mathcal{B}$, the resonance peak in the $\omega_s$-dependence of the transmission becomes sharper and the linewidth of the amplification peak tends to zero as the oscillation threshold is approached~\footnote{Simple algebraic manipulations show that $E_t(\omega_s)$ and $E_{pc}(\omega_s)$ exhibit a pair of resonant transmission poles as a function of $\omega_s$. In the simplest case $\bar{\Omega}=2\omega_c$, the two poles are at $\omega_c-i(\gamma_c/2-2\mathcal{B})$  and $\omega_c-i(\gamma_c/2+2\mathcal{B})$. In particular, note how the imaginary part of the first pole tends to zero when approaching the parametric threshold $\mathcal{B}\rightarrow \gamma_c/4$. }

A similar $\omega_s$-dependence is apparent in the amplitude $E_{pc}$ of the phase-conjugated beam that emerges from the cavity at  the specular frequency $\bar{\Omega}-\omega_s$  [Fig.~\ref{fig:stimulated}(b)]. The amplitude of this latter beam vanishes in the absence of any modulation, so that its very presence is a consequence of a dynamical Casimir mixing process of the incident field by the temporal modulation of the refractive index. 

For $\bar{\Omega}=2\omega_c$, the $\omega_s$-dependence is also peaked at $\omega_c$ and shows a linewidth narrowing phenomenon as the threshold is approached $\mathcal{B}\rightarrow \gamma_e/4$. 
In the presence of a finite detuning of the modulation $\bar{\Omega}-2\omega_c\neq 0$, the transmission peak splits at low $\mathcal{B}$ into a doublet, one peak being close to $\omega_c$, the other peak being close to $\bar{\Omega}-\omega_c$.
As $\mathcal{B}$ is increased, the peaks move closer to each other and finally merge. As the threshold is approached, the splitting transfers into the imaginary parts as in the $\bar{\Omega}=2\omega_c$ case.

We conclude this section with some quantitative remarks about the actual intensity of the emitted DCE radiation. To this purpose, we insert realistic parameters into the expression \eq{N} for the photon flux emerging from the cavity. From the definition of $\mathcal{B}$ in \eq{Afin}, it is immediate to see that the result is dominated by the Gaussian $\exp(-\bar{\Omega}^2\bar{\tau}^2/4)$ and a critical condition to have an appreciable photon generation is that that $\bar{\tau} \lesssim 2/\omega_c = \tau_c/\pi$: the effective duration $\bar{\tau}$ of the Gaussian pump pulses in the cavity has to be comparable or shorter than the optical period $\tau_c$ of the cavity mode. Otherwise, the cavity mode sees the modulation of the index as almost adiabatic and remains in the ground state with no photons. On this basis, it can be advantageous to tune the repetition rate close to twice the cavity frequency, $\Omega\simeq 2\omega_c$, so to work with the fundamental component of $\mathcal{A}(\omega)$ for which the Gaussian factor is least crucial.

Moreover, for a given value of $\delta n^{\rm peak}$ and a fixed $\tau/T$, the $\omega_c$ factor in \eq{A0} suggests it is advantageous to shift the cavity frequency $\omega_c$ (and correspondingly the repetition rate $\Omega$) towards shorter wavelengths, e.g. in the visible domain. Indeed the number of cavity photons is predicted (in the lossless case described by Eq.~(\ref{gamma}) and considered in the numerical calculations of the next section) to grow exponentially with $\mathcal{B}\propto\omega_c$.  However, there is a trade-off between increasing the frequency and the actual possibility to achieve the necessary modulation rates. Trains of $\sim1$ fs pulses with a few micron periodicity have been demonstrated and sub-fs durations with 1 micron periodicity should be relatively easily generated as discussed above. However, shorter periodicities would require shifting the spectrum further into the UV region, thus encountering issues with material absorption and a general difficulty in obtaining high energy pulses e.g. at the fourth harmonic of an 800 nm laser pulse. We therefore believe that, whilst the laser source scheme adopted here is perfectly reasonable, it does also seem to be the limit allowed by current technology.

Realistic values for the pump train of pulses can be summarized as follows: combining the first four harmonics of the fundamental wavelength at $\Lambda=1\,\mu$m, a pulse duration  on the order $\tau=1$~fs can be obtained. Realistic values for the cavity parameters can be $\delta n^{\rm peak}=0.003$, $\sigma_p/L_c=0.5$,  $\gamma_c=6\cdot10^{11}\,\textrm{s}^{-1}$ and $\sigma_c=0.5$ $\mu$m, which leads to the quite optimistic prediction $\Phi_{\rm DCE}\sim 10^{10}\,\textrm{s}^{-1}$ for the emitted photon flux\footnote{Note that for a propagation speed $v_p\leq c$, a pulse duration of 1~fs corresponds to a length $\sigma\leq 0.33\,\mu$m. An important contribution to $\bar{\tau}$ then comes from the mode waist $\sigma_c$. }. Assuming that the laser pulse train lasts for 1 ns, we predict a  DCE photon emission rate of the order of $10$ photons/pulse that holds strong promise for detection.  And it is worth noting the value of $\mathcal{B}$ is also not far below the threshold value $\mathcal{B}_{\rm thr}$ for DCE parametric oscillation.

\section{Numerical calculations}
\begin{figure}[!t]
\centering
\includegraphics[width=8cm]{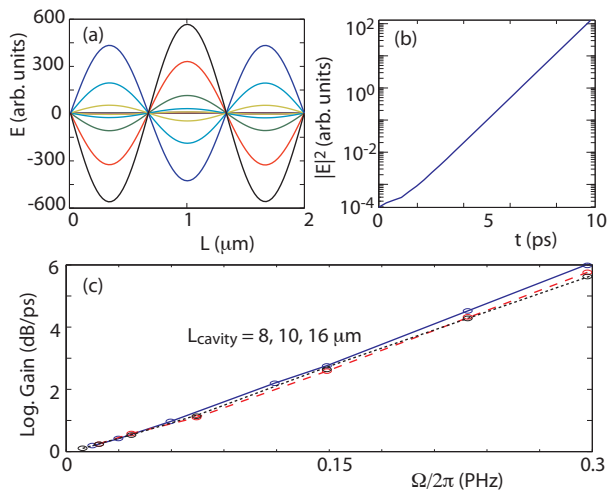}
\caption{\label{fig:gain}  Numerical simulations for a cavity excited exactly on resonance: $\Omega=2\omega_c=(2\pi) 0.3$ PHz, $\delta n=0.003$. Snapshots (a) of the electric field inside the cavity for a series of increasing times and field intensity versus time (b) for a cavity of length $L_c=2\,\mu$m.  (c) Logarithmic gain versus modulation frequency $\Omega$ for three different cavity lengths, 8 $\mu$m (solid line), 10 $\mu$m (dashed line), 16 $\mu$m (dotted line).  The modulation is always taken on resonance with a cavity mode, $\Omega=2\omega_c$.}
\end{figure}

The calculations of the previous section were based on a quantum optical description of the evolution of the field amplitude in a single cavity mode coupled to a dissipative bath of external radiation modes. A most significant advantage of this approach is the possibility of including non-linearities in the model, so to describe the back-action effect of the DCE onto the external modulation~\cite{kardar,noi_back}. 

In this section we explore a different strategy to theoretically study the DCE, based on the FDTD solution of the classical Maxwell equations with a time-dependent refractive index. Zero-point noise in the initial quantum vacuum state is included in a phenomenological way via a noisy initial condition. A key advantage of this approach is that it is able to follow the system evolution in real time and therefore quantify the actual time-scale over which the DCE gain can be observed. Another promising point is the possibility of extending the calculation to cavities of arbitrary geometry, so to fully include effects stemming from the multi-mode nature of the field.
\begin{figure}[!t]
\centering
\includegraphics[width=8cm]{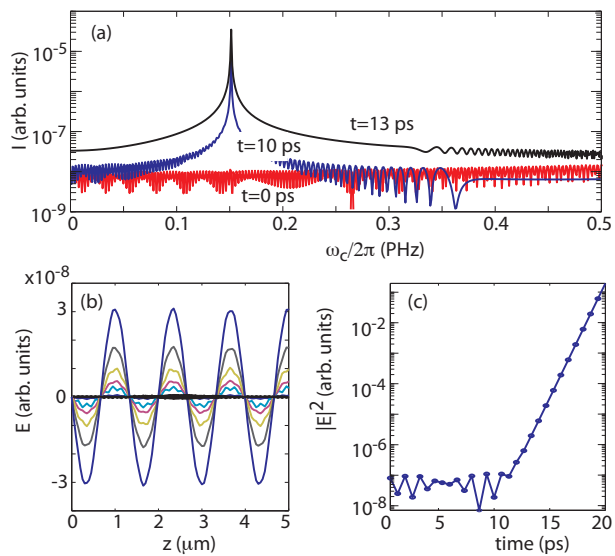}
\caption{\label{fig:noise}  Numerical simulations of the DCE in a cavity of length $L_c=80$ $\mu$m  seeded by noise.  The modulation at $\Omega=(2\pi)\,0.3$ PHz is resonant with a cavity mode and has $\delta n=0.003$. (a) input spectrum at t=0~ps (red line) and spectra at two different times, t = 10 (blue line) and 13 ps (black line). (b) Electric field distribution inside the cavity (zoomed in to the first 5 $\mu$m) for six different successive times starting at t=0 ps (black line). (c) Temporal evolution of the field intensity inside the cavity. }
\end{figure}

We start by considering the simplest geometry of a spatially homogeneous cavity material  of refractive index $n_0=1.5$ enclosed between perfectly reflecting cavity mirrors. These are simulated by imposing perfectly reflecting boundary conditions to the field \cite{yee}.  The cavity index is made to sinusoidally  oscillate with an amplitude $\delta n$ and a seed pulse is inserted as an initial condition. Figure~\ref{fig:gain} shows an example for $\delta n=0.003$ and a $L_c=2\,\mu$m long cavity  with a seed pulse wavelength resonant with the cavity mode that has a wavelength (in vacuum) $\lambda_c=2\,\mu$m and excited under the resonance condition $\Omega=2\omega_c$. In Fig.~\ref{fig:gain}(a) we show the increasing electric field amplitude for different times and  in Fig.~\ref{fig:gain}(b) we show the field intensity inside the cavity as a function of time. 
In Fig.~\ref{fig:gain}(c) we show the logarithmic gain, 
\begin{equation}
G=10\,\frac{d}{dt}\log_{10}\left[\frac{|E(t)|^2}{|E(0)|^2}\right],
\end{equation}
 for three different cavity lengths $L_{c}$ and varying modulation frequency $\Omega$: as expected from the model, the gain does not depend on $L_{c}$. On the other hand,  as $\Omega$ is increased, a linear increase of the gain, implying an exponential increase of the  field intensity inside the cavity  is observed, in agreement with the analytical calculations. 

In Fig.~\ref{fig:noise} we show another simulation in which we insert noise as an input condition: this phenomenological way of including in a classical FDTD simulation the zero-point fluctuations of the field in its quantum vacuum state can be made rigorous in terms of quantum Langevin equations or using the Wigner representations of the quantum field \cite{wallsmilburn,quantumnoise}. 
Figure~\ref{fig:noise} shows the initial spectrum (t=0 ps) and the spectrum after 10 and 13 ps (a), and the electric field distribution at different times (b - the thin black line shows the initial ``noise'' condition) with $\delta n=0.003$ and $\Lambda=2\pi c/\Omega=1$ $\mu$m. 
A single peak is clearly amplified at the resonance condition $\omega_c=\Omega/2=(2\pi)0.15$ PHz. The fact that that a steady exponential growth sets in after just $\sim10$ ps, as seen in Fig.~\ref{fig:noise}(c) is an important indication of the accuracy of the analytical single-mode model.

\begin{figure}[!t]
\centering
\includegraphics[width=8cm]{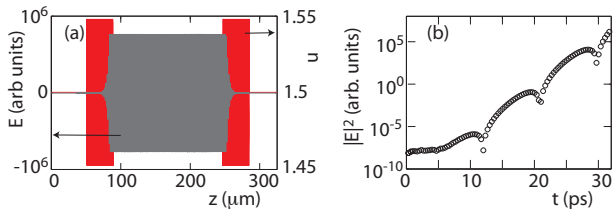}
\caption{\label{fig:grating} Numerical simulations of the DCE in Bragg grating cavity. (a) grating index profile and electric field amplitude at t=30 ps. The cavity index is modulated at $\Omega=(2\pi)\,0.3$ PHz and $\delta n=0.003$. (b) The field intensity evolution with time inside the cavity.}
\end{figure}

Finally, in Fig.~\ref{fig:grating} we show a simulation in which we substitute the perfect cavity mirrors used above with Bragg grating mirrors: the cavity is thus composed of two 30 layer Bragg reflectors (sinusoidal refractive index variation between 1.45 and 1.55) that enclose a 170 $\mu$m long cavity that has $n_0=1.5$ and $\delta n=0.003$. The cavity length was fine tuned so as to exhibit a resonance at 2 $\mu$m wavelength. The cavity and the electric field distribution at the end of the simulation are shown in Fig.~\ref{fig:grating}(a). The intensity evolution inside the cavity modulated at resonance ($\Lambda=1\,\mu$m) is shown in logarithmic scale in Fig.~\ref{fig:grating}(b): after $\sim 10$ ps a steady growth regime is reached with a rate of $\sim4$~dB/ps. The oscillations in the field intensity appear to be related to the beating of the modulation with the propagation of the pulse back and forth in the cavity with a periodicity that depends on the cavity length. With very short cavities this effect is not observable, but longer cavities lead to this characteristic beating with a periodic cavity modulation. Similar behavior is obtained with a wide range of input parameters (e.g. cavity lengths, $\delta n$).

\section{Conclusions}
In conclusion, we have proposed and quantitatively characterized an experimental scheme for measuring the dynamical Casimir effect: the effective optical length of a cavity is varied in time on a PHz time scale by means of a train of ultrashort pulses that modulate the effective refractive index of the cavity material. The predicted rate of photon emission for realistic pulse train and cavity parameters appears promising in view of experimental studies. The over-arching idea of laser pulse-induced cavity modulation is very general and may be extended to other settings such as acoustic cavities modulated by laser-pulse trains \cite{vahala}.

\acknowledgments
IC acknowledges financial support from ERC through the QGBE grant. The authors acknowledge discussions with M. Kolesik and E. Abraham.

\end{document}